\documentclass[12pt]{iopart}

\begin{document}

\title{Generalized 
%Kustaanheimo—Stiefel 
KS transformations,  $ND$ singular oscillator and  generalized MICZ-Kepler system}

\author{A. Lavrenov}

\address{Belarusian  State Pedagogical  University,  Minsk, Belarus, 
18, Sovetskaya St., Minsk, Belarus, 220050}
\ead{lavrenov@bspu.by}
\vspace{10pt}
\begin{indented}
\item[]May 2019
\end{indented}

\begin{abstract}
The description of number of dual (quasy)-exactly solvable models with its hidden symmetry algebra has been given at different levels of analysis within the framework of generalized Kustaanheimo-Stiefel (KS)-transformations. It's shown that $N$-dimensionall singular oscillator and $(n+1)$-dimensional generalized MICZ-Kepler system are dual to each other and the duality transformation is the generalized version of the KS transformation. The solvability of the Schr\"{o}dinger equation of these problems by the variables separation method is given in a double, spherical and parabolic coordinates. The quadratic Hahn algebra $QH(3)$ as a hidden symmetry remains unchanged with different decomposition of the original real space ${\rm I \!R}^{N}$ into components in the framework of addition rule for $SU(1,1)$ algebra. Also the hidden symmetry algebra as Higgs/Hahn algebras is clearly shown by the commutant approach in the sense of Howe duality.  A dimensional reduction is carried out to a singular oscillator of two dimensions where every variable is the $n$-dimensional hyper-radius $r$. The dual connection with $N \equiv 2n$-dimensional singular oscillator and the $(n+1)$ general MICZ-Kepler system in the class of quasi-exact (QE) problems also are considered. The certain generalization both of harmonic  oscillator model by anisotropic and nonlinear inharmonic terms and its dual analog is shown and analyzed in the framework of generalized KS transformations. The exact analytical solutions of the Schr\"{o}dinger equation for abovementioned problems for QE class are discused and given for four series of dual quasi-exact solvable models. In particular, a comparison with similar results in lower dimensions and its generalization are given.

\end{abstract}

%
% Uncomment for keywords
%\vspace{2pc}
%\noindent{\it Keywords}: XXXXXX, YYYYYYYY, ZZZZZZZZZ
%
% Uncomment for Submitted to journal title message
%\submitto{\JPA}
%
% Uncomment if a separate title page is required
%\maketitle
% 
% For two-column output uncomment the next line and choose [10pt] rather than [12pt] in the \documentclass declaration
%\ioptwocol
%

\section{Introduction}

There is the Kustaanheimo and Stiefel transformation  (KS transformation) \cite{01} that establishes  a connection between the problem of an isotropic harmonic oscillator (HO) in a real four-dimensional space and the Kepler problem in ordinary three-dimensional space (see \cite{02}-\cite{04}). 
On the other hand, it's possible to establish the connection between the Coulomb problem and the harmonic oscillator in the dimensions of its spaces, realizing the Hopf fibrations through the generalized KS transformations
The Hopf maps are fibrations of spheres over spheres with the fiber-sphere \cite{hopf1931}. The term "Hopf fibration or map" is usually restricted to the fibrations between spheres  $S^{2n-1}/S^{n-1}=S^n$. Among them, four Hopf maps stand out: $S^{2n-1}/S^{n-1}=S^n$, ($n=1,2,4,8$):

\begin{itemize}
 \item  $S^{1} \to  S^{1}$   with fiber $S^{0}$,  (zero Hopf map $n=1$),
 \item  $S^{3} \to  S^{2}$  with fiber $S^{1}$ (first Hopf map $n=2$),
 \item $S^{7} \to  S^{4}$   with fiber $S^{3}$ (second Hopf map $n=4$)
  \item $S^{15} \to  S^{8}$   with fiber $S^{7}$  (third Hopf map $n=8$)
\end{itemize}

Each such Hopf map is closely related with it's division algebra:  $n=1$ reflects the existence of division algebra of real numbers, $n=2$ - complex numbers,  $n=4$ – quaternions and $n=8$ - octonions.  Zero Hopf map is related with anyons (or magnetic  vortices) \cite{ntt}. The Dirac and Yang monopoles are related with  the first and the second ones  respectively.
 The first three cases have been studied in detail for a long time. There are a lot of literature that is devoted to this topic, which has already developed academic logic and style of presentation of the materials for this topic (for example \cite{bel2} - \cite{ple4}). The authors \cite{van1} - \cite{van6} have thoroughly worked out many questions of the above  last case of the Hopf bundle associated with 16-dimensional harmonic oscillator and 9-dimensional Coulomb problems recently.
All these connections between the Coulomb problem and the harmonic oscillator are very helpful for solutions of many problems of atomic physics (see \cite{lik30}-\cite{lik32}). 

However, in his  \cite{lik33}, Cordani, using the methods of group theory, asserted that one can construct the required KS transformations for the Kepler problem with dimension $n > 6$, but  didn't answer the question ''how these transformations can be realized?''.  The given theory was built in the article \cite{van0}. The authors of this article gave a simple graphical algorithm for constructing the generalized KS transformations that realizes the connection between two problems above and proved the following proposition.

\textbf {Proposition.} \textit{ A connection between the Kepler problem in a real space of dimension $n+1$ and the problem of an isotropic harmonic oscillator in a real space of dimension $N$ exists and can be established by means of generalized KS transformations in the cases when $N=2n$ and $n=2^h (h=0, 1, 2, \ldots)$. }

The question naturally arises - is it possible to obtain results realized in small dimensions for the general case? 

From our point of view, the answer is YES.
Therefore, in the framework of generalized KS transformationsthat that is described above, we aim to describe a series of dual exactly solvable models with their hidden symmetry algebra at different levels of analysis.
The article is organized as follows. In Section II,  we give a brief results of the works by the authors of \cite{van0}, recall the addition rule for $SU(1,1)$ algebra, the problem of hidden symmetry for the systems admitting such an approach in accordaning to Ref. \cite{Zhe1} and the well-known results on Heisenberg-Weyl, $SU(1,1)$, $U(16)$, Higgs and Hahn algebras. 

Our main result will be the object of Sections  III - V where the dual connection with the general MICZ-Kepler system and $N$-dimensional singular  oscillator will be made; the exact analytical solutions of the Schr\"{o}dinger equation for these  problems will be successfully built  in a double, spherical and parabolic coordinates; the  hidden symmetry algebra as Higgs/Hahn algebras will be clearly shown by both addition rule for $SU(1,1)$ algebra and the commutant approach in the sense of Howe duality. 
In Section VI, a dimensional reduction is carried out to a singular oscillator of two dimensions, each variable of which is the $n$-dimensional hyper-radius $r$.
The dual connection with $N \equiv 2n$-dimensional singular  oscillator and the $(n+1)$  general MICZ-Kepler system in  the class of  quasi-exact problems  are considered in the last sections VII.

\section{Generalized KS transformations, the famous algebras  and the addition rule for $SU(1,1)$ algebra}

The problem of an isotropic harmonic oscillator (IHO) in a real $N$-dimensional space is described by  the next Schr\"{o}dinger equation:
%=====================================================================
\begin{eqnarray}
   \left(
   H_0 - 4Z
\right)
  \psi (u) \equiv
  \left(-\frac{1}{2}
\frac{{\partial}^2}{\partial u_{s} \partial u_{s}} +
\frac{1}{2}
%\frac{ \omega^2}{2}
%\left( 
\omega^2 u_{s}  u_{s} - 4Z
\right) \psi (u) = 0,
\label{1}
\end{eqnarray}
%=====================================================================
Here, 
%$u_s (s=1, \ldots, N)$ 
$u_s (s=\overline {1, N})$
are the Cartesian coordinates of the point $u$ in $N$-dimensional space; summation over repeated indices is understood; $w$ is a real positive number; $Z$ is the eigenvalue. A generalized KS transformation  is determined by 
%=====================================================================
\begin{equation}
\eqalign{
%\begin{eqnarray}
x_{\lambda}  =(\Gamma_{\lambda})_{st} u_{s} u_{t},  \  & (
\overline {\lambda = 1, n + 1)}, \nonumber\\
%\lambda = 1, \ldots, q + 1), \\
\chi_{a}=\chi_{a}(u) & 
(a=\overline {n+2, N \equiv 2n }),
%(a=q+2, \ldots, N),
\label{2}
}
\end{equation}
%\end{eqnarray}
where the real symmetric $N \times N$ matrices $\Gamma_{\lambda}$ satisfy 
%\begin{eqnarray}
\begin{equation}
\eqalign{
Sp \  (\Gamma_{\lambda}) = 0,  \quad \Gamma_{\lambda}\Gamma_{\mu} + \Gamma_{\mu}\Gamma_{\lambda} = 2\delta_{\lambda \mu}I;
%\label{3} \\
\nonumber\\
x_{\lambda} x_{\lambda} = (u_s u_s)^2 = r^2  \quad (r = u_s u_s);
\label{3}
%\end{eqnarray}
}
\end{equation}
$I$ is the unit $N \times N$ matrix and  $\chi_{a} (u)$  are any functions with  nonzero Jacobian of the transformation $\chi$ from the variables $u$ to the variables $x$. 
In according to \cite{van0} among the solutions of Eq. (\ref{1})  it is possible to find functions $\Psi$ that are dependent on the variables $x$ only and satisfied the equation  of the Kepler problem in a space of dimension $n+1$.
%=====================================================================
\begin{eqnarray}
   \left(
   H'_0 - E
\right)
\Psi (r)
 \equiv
  \left(-\frac{1}{2}
\frac{{\partial}^2}{\partial x_{\lambda} \partial x_{\lambda}}
-\frac{Z}{r} +
\frac{1}{8}
%\frac{ \omega^2}{2}
%\left( 
\omega^2  
%-Z
\right)  \Psi (r) = 0,
\label{4}
\end{eqnarray}

On the other hand, with the help of the pair of creation (birth)
 $a_s^{\dagger}=\sqrt\frac{\omega}{{2}}\left(-\frac{1}{\omega}\frac{\partial}{\partial u_s}+u_s\right)$
and annihilation (destruction)  operators
$ a_s=\sqrt\frac{\omega}{{2}}\left(\frac{1}{\omega}\frac{\partial}{\partial u_s}+u_s\right),$
 the  number operators
$N_s= a_s^{\dagger}a_s \equiv \left[ -\frac{1}{2 \omega}\frac{\partial^{2}}{\partial {u_s}^{2}}+ \frac {\omega}{2}{u_s}^{2}- \frac {1}{2} \right]$
the Hamiltonian of the IHO can be rewritten in the operator form:
\begin{eqnarray}\label{5}
 H_0=\omega\left(\sum\limits_{s=1}^{2n} N_s+n\right)
\end{eqnarray}
Recall the definition of the Heisenberg-Weyl algebra $W(n)$ and following commutation relations between  its operators:
\begin{eqnarray*}
[a_i; a^{\dagger}_{j} ]=\delta_{ij},    \qquad  
 [N_i,a_j] =-a_i\delta_{ij}, \qquad
 [N_i,a_j^{\dagger}]=a_i^{\dagger}\delta_{ij}.  \qquad 
%  i,j=1,\dots,n.
   i,j=\overline {1, n}
\end{eqnarray*}
The Lie algebra $U(n)$ with generators $E_{ij}$, $ i,j=\overline {1, n}$ admits the following realization in $W(n)$:
\begin{eqnarray*}
 E_{ij}=a_i^{\dagger}a_j,\qquad   i,j=\overline {1, n}.
\end{eqnarray*}
 Hamiltonian  $H_0$
commutes with both $E_{ij}$ (i.e. $[H_0,a_i^{\dagger}a_j]=0$) and with generators of infinitesimal rotations
\begin{eqnarray*}
 L_{jk}=\frac{i}{2}(a_ja_k^{\dagger}-a_j^{\dagger}a_k)=-\frac{i}{2}\left(u_j\frac{\partial}{\partial u_k}-u_k\frac{\partial}{\partial u_j}\right)
\end{eqnarray*}
which have well known  commutation relations
\begin{eqnarray*}
 [L_{jk},L_{\ell m}]=\frac{i}{2}(L_{j\ell}\delta_{km}-L_{k\ell}\delta_{jm}+L_{km}\delta_{j\ell}-L_{jm}\delta_{k\ell}),\qquad j,k,\ell,m=\overline {1, n}.
\end{eqnarray*}

The metaplectic representation of $SU(1,1)$ is defined by the following map in $W(1)$:
\begin{eqnarray}\label{6}
 J_0^{(i)}=\frac{1}{2}\left(N_i+\frac{1}{2}\right),\qquad J_+^{(i)}=\frac{1}{2}a_i^{\dagger}{}^{2},\qquad J_-^{(i)}=\frac{1}{2}{a_i}^{2},
\end{eqnarray}
where the generators  $J_0$, $J_{\pm}$  obey the commutation relations
\begin{eqnarray*}
 [J_0,J_\pm]=\pm J_\pm,\qquad [J_+,J_-]=-2J_0.
\end{eqnarray*}
Its Casimir element is given by
\begin{eqnarray*}
 Q={J_0}^{2}-J_+J_--J_0.
\end{eqnarray*}
Consider now the addition of two irreducible representations of $SU(1,1)$ for which the initial Casimir operators take values~  $Q^{(i)}=\lambda_i(\lambda_i- 1)$,~  $i=\overline {1,2}$, and let us write
\begin{eqnarray}
 J_0^{(12)}=J_0^{(1)}+J_0^{(2)}, \qquad J_\pm^{(12)}=J_\pm^{(1)}+J_\pm^{(2)}
\label{7}
\end{eqnarray}
with the superindex denoting on which of the two factors in $SU(1,1)^{\otimes 2}$ the operator is acting.
It is obvious that Hamiltonian $H= J_0^{(12)}$  has dynamical symmetry as $SU(1,1) \oplus SU(1,1)$ and it's spectrum  $\epsilon=n_1+n_2+  \lambda_{1}+ \lambda_{2}$ is degenerated. This means that there are the integrals of motion commuting with $H$.
A trivial example  of such integrals for the possible combinations from initial operators $(J_0^{(1)}, J_{\pm}^{(1)}, J_0^{(2)}, J_{\pm}^{(2)})$ is as follows:
\begin{equation}
\eqalign{
%\begin{eqnarray}
K_1 = J_0^{(1)} - J_0^{(2)} \nonumber\\
K_2= Q^{(12)} = Q^{(1)}+ Q^{(2)} + 2J_0^{(1)} J_0^{(2)} -
J_+^{(1)} J_-^{(2)}-
J_-^{(1)} J_+^{(2)}
\label{8}
}
%\end{eqnarray}
\end{equation}
The hidden symmetry of our  Hamiltonian $H= J_0^{(12)}$  means  the existence of some  algebra constucted from the integrals $K_1, K_2$.  According to the work \cite{Zhe1} such algebra is quadratic Hahn algebra  $QH(3)$ under the commutation relations:
\begin{equation}
\eqalign{
%\begin{eqnarray}
[K_{1}, K_{2}] = K_{3};  \nonumber\\
\left[
K_{2}, K_{3}
\right] =  -2(K_{1} K_{2}+K_{2} K_{1}) + \delta_1;   \nonumber\\
%[K_{2}, [K_{1}, K_{2}] ]=
%& 2 K_{2}K_{1}K_{2} - K_{2}^2K_{1}-K_{1}K_{2}^2  \nonumber\\
%4\epsilon (Q^{(1)}-Q^{(2)});\\
%\label{10}
\left[
K_{3}, K_{1}
\right] =  -2K_{1}^2-4K_{2} +  \delta_2
%& 2 K_{1}K_{2}K_{1} - K_{1}^2K_{2}-K_{2}K_{1}^2 \nonumber\\
%2 \epsilon^2 +4\epsilon (Q^{(1)}+Q^{(2)}) 
\label{9}
%\end{eqnarray}
}
\end{equation}
Recall that
the Higgs algebra can be considered as a polynomial deformation of $ SU(2) $ with three generators $D$, $A_{\pm}$ satisfying the following commutation relations  \cite{H6} - \cite{L7}:
\begin{eqnarray}
[D, A_{\pm}]= \pm 4 A_{\pm}; \qquad
%\nonumber\\
[A_{+}, A_{-}]= -D^{3}+\alpha_{1} D+\alpha_{2} 
\label{10}
\end{eqnarray}
with $\alpha_1$, $\alpha_2$ central elements.
According to  \cite{Zhe13},
the isomorphizm of the discrete version of the Hahn algebra and the Higgs algebra  is readily seen by taking
\begin{equation}
\eqalign{
%\begin{eqnarray}
\label{11}
 K_1=\frac{1}{2}D  \nonumber\\
 K_2=-\frac{1}{4}\left(A_++A_-+\frac{1}{2}D^{2}\right)+\frac{\alpha_1}{8}  
%\end{eqnarray}
}
\end{equation}
and observing that the commutation relations (\ref{9}) then follow from (\ref{10}) with
\begin{equation}
\eqalign{
%\begin{eqnarray}
\label{12}
 \delta_1 \equiv -\frac{\alpha_2}{4}  \equiv   4\left(J_0^{(1)}+J_0^{(2)}\right)\left(Q^{(1)}-Q^{(2)}\right)  \nonumber\\
 \delta_2 \equiv \frac{\alpha_1}{2}    \equiv    2\left(J_0^{(1)}+J_0^{(2)}\right)^{2}+4\left(Q^{(1)}+Q^{(2)}\right). 
%\end{eqnarray}
}
\end{equation}

\section{Dual connection between  the general MICZ-Kepler system and $N$-dimensional singular  oscillator}
In this section, we propose to discuss the following questions: How does the structure of matrices for the generalized version of the
KS transformation impose restrictions on the splitting of the initial problem for the IHO into its components? Which version of the generalization for the IHO components  choose? What will be its dual partner for the chosen option of generalization? Are the proposed dual models exactly solvable?

So, a generalized KS transformation  is based on the matrices  $\Gamma_{\lambda}$.  
 The graphical algorithm for their construction, proposed in \cite{van0}, is characterized by repetition of the structure of old matrices of dimension $k$ in the structure of a new matrix of double dimension $2k$ on the it's main diagonal. In particular, the choice in \cite{van0}  was such
 for  the matrix $\Gamma_{n+1}$
\begin{displaymath}
\Gamma_{n+1}=
\left( 
\begin{array}{cc}
 I & 0 \\
0 & - I  \\
\end{array} 
\right)
\end{displaymath}
where $0$ and $I$ are the null and unit $n \times n$ matrices, i.e.
%  for $x_{n+1}=(\Gamma_{n+1})_{st} u_{s} u_{t}$
\begin{equation}
\eqalign{
%\begin{eqnarray}
x_{n+1}= (\Gamma_{n+1})_{st} u_{s} u_{t} \nonumber\\
\qquad =u_{1}^2+u_{2}^2+ \ldots + u_{n}^2 - u_{n+1}^2 - u_{n+2}^2 - \ldots u_{2n}^2 
% \nonumber\\
%&\equiv &u_{1}^2+u_{2}^2+ \ldots + u_{n}^2 - v_{1}^2 - v_{2}^2 - \ldots v_{n}^2   \nonumber\\
%&\equiv &u_{i}  u_{i} - v_{i}  v_{i}; \nonumber\\
%r&= &\sqrt{ x_{\lambda} x_{\lambda}} = u_{i}  u_{i} + v_{i}  v_{i}; \
\label{13}
%\end{eqnarray}
}
\end{equation}
Then the remaining matrices
$\Gamma_{\lambda}, (\lambda=\overline {  1, n})$ 
% 1, \ldots, q)$ 
 necessarily have the structure
\begin{displaymath}
\Gamma_{\lambda}=
\left( 
\begin{array}{cc}
0 & \gamma_{\lambda} \\
\gamma_{\lambda} & 0  \\
\end{array} 
\right)
\end{displaymath}
where $\gamma_{\lambda}$ are anticommuting real $n \times n$ matrices each of whose square is equal to the unit matrix. 

Facts that have been described above  and the results obtained in small dimensions clearly indicate the need to consider the original problem of an IHO in a real $N \equiv 2n$-dimensional space as the sum of the two ones of dimension $n$:
%=====================================================================
\begin{eqnarray}
%\hat 
{ H_{0}}  \psi({\bf u, v})=
 \left[-\frac{1}{2}(
\frac{{\partial}^2}{\partial u_{i} \partial u_{i}} +
\frac{{\partial}^2}{\partial v_{i} \partial v_{i}}) +
\frac{ \omega^2}{2}
\left( u_{i}  u_{i}+v_{i}  v_{i}
\right)\right] \psi =4 Z  \psi,
\label{14}
\end{eqnarray}
%=====================================================================
Here  $ \omega, Z$  have positive real values and are correspondingly the frequency and energy of the IHO, $v_i \equiv u_{n+i}, \ (i=\overline {1, n})$.

To preserve symmetry between two IHO of dimension $n$ and also taking into account the results obtained in small dimensions, we choose the same perturbation for both of them, which will be additive and the simplest.
Therefore
we proposed next simple generalization of our initial problem to  so-called a double singular oscillator:
%===================================================================
\begin{eqnarray}
%\hat
 {H} \ \psi({\bf u, v})=
\left[{
%\hat
 H}_0+
\frac{c_1}{u^2_1+...+u^2_n}+
\frac{c_2}{v^2_{1}+...+v^2_n}
\right] \psi = Z  \psi,
\label{15}
\end{eqnarray}
%=====================================================================
where ${
%\hat
 H}_0$ is Hamiltonian  of an IHO determined earlier in the explicit formula for (\ref{1}).

Thus, there is a transparent transition for any generalizations with additive terms to the original Hamiltonians.
For example,  it takes place
\begin{equation}
\eqalign{
%\begin{eqnarray}
x_{n+1}=u_{1}^2+u_{2}^2+ \ldots + u_{n}^2 - v_{1}^2 - v_{2}^2 - \ldots - v_{n}^2   \nonumber\\
\qquad \equiv u_{i}  u_{i} - v_{i}  v_{i}; \nonumber\\
\ r \quad = \sqrt{ x_{\lambda} x_{\lambda}} = u_{i}  u_{i} + v_{i}  v_{i}; \
\label{16}
%\end{eqnarray}
}
\end{equation}
or
  %================================================================
\begin{eqnarray*}
\frac{c_1}{u^2_1+...+u^2_n}+
\frac{c_2}{v^2_{1}+...+v^2_n} = \frac{2 c_1}{2 u_{i}  u_{i}}+
\frac{2 c_2}{2 v_{s}  v_{s}}= \\
=\frac{2 c_1}{r+x_{n+1}} +
\frac{2 c_2}{r-x_{n+1}}=
\frac{4 \lambda_1}{r+x_{n+1}} +
\frac{4 \lambda_2}{r-x_{n+1}}
\end{eqnarray*}
%==================================================================

This fact means that the dual analogue of so-called a double singular oscillator is the generalized MICZ-Kepler system with non central terms:
  %================================================================
\begin{eqnarray}
%\hat
 {H'} \ \psi({\bf r})=
\left[
%\hat 
{ H'_{0}}+
\frac{\lambda_1}{r(r+x_{n+1})} +
\frac{\lambda_2}{r(r-x_{n+1})}
\right)\psi= E \psi
,\label{17} 
\end{eqnarray}
%==================================================================
where $ 2\lambda_a = c_a, a=1,2$ are nonnegative constants; ${
%\hat
 H'}_0$ is Hamiltonian  of the Kepler problem determined earlier  (\ref{4}).

Their dual nature underlines the fact that the roles of $ E = -\frac{\omega^2}{8}$ and $Z$  are interchanged. The variables $E$ and $Z$ become  a negative number that denotes the energy of bound states and a parameter defining the `charge'  value in the Coulomb potential respectively. 

So, on the one hand, our proposed generalization in according to the above formulas is reduced to additive terms in the Hamiltonians, and on the other hand we get a rich symmetry picture of new problem, which leads to a series of exactly-solvable cases.

\subsection{Variables separation  in double coordinates}

The Schr\"{o}dinger equation for a $N \equiv 2n$ -dimensional double singular oscillator ($\ref{15}$)  represents the sum of two singular oscillators of dimensions $D=n$:
\begin{equation}
\eqalign{
%\begin{eqnarray}
H=H_1+H_2=\sum_{1\le a\le 2}H_a=\nonumber\\
\quad =
\sum_{1\le a\le 2}
\left[
-\frac{1}{2}
\frac{\partial^2}{\partial x^a_{i} \partial x^a_{i}} +
\frac{ \omega^2}{2} ( x^a_{i} x^a_{i}) +
\frac{c_a}{x^a_{i} x^a_{i}}
\right],
\label{18}
%\end{eqnarray}
}
\end{equation}
where $x^1_{i} =u_{i}, \quad x^2_{i} =v_{i}$.

In each $nD$ real space ${\rm I \!R}^n$ where   $ x_{1}, x_{2},\ldots,x_{n}$ are Cartesian coordinates, we introduce  it's hyperspherical coordinates: ($\phi_{1},\dots, \phi_{n-1}$)  are the hyperspherical angles and $r$ is the hyperradius by
%=================================
\begin{equation}
\eqalign{
%\begin{eqnarray}
%\nonumber  
x_{n} \quad =r\cos(\phi_{n-1}), \nonumber\\ 
x_{n-1}=r\sin(\phi_{n-1})\cos(\phi_{n-2}),\nonumber\\
\qquad ...\nonumber\\
\qquad ...\nonumber\\ 
x_{2} \quad =r \sin(\phi_{n-1})\sin(\phi_{n-2})\cdots \sin(\phi_{2}) \cos(\phi_{1}),\nonumber\\ 
x_{1} \quad =r \sin(\phi_{n-1})\sin(\phi_{n-2})\cdots \sin(\phi_{2}) \sin(\phi_{1}),
\label{19}
%\end{eqnarray}
}
\end{equation}
%=====================================================================

Using the ansatz $ \Psi({\bf r})=\Psi(r,\phi_{1},\cdots,\phi_{n-1})=R(r) \Omega(\phi)$
the Schr\"{o}dinger equation for  each $H_a$ can be rewritten in terms of it's separated equations of hyperradius $r$ and angular variables $\phi$ as follows:
%==================================================
\begin{eqnarray}
\left[
\frac{1}{r^{n-1}}\frac{\partial}{\partial r} ( r^{n-1} \frac{\partial}{ \partial r})+
\left(
2 E_a-\omega^2 r^2
\right)-
\frac{\Lambda^2+2c_a}{r^2}
\right]R (r)=0,\label{20} \\
\left[
\Lambda^2-L(L+n-2)
\right]
\Omega(\phi)=0,
\label{21}
\end{eqnarray}
%==================================================
where $E_a $ are the eigenvalues of $H_a$, but $E_1+E_2=4Z;  \quad L(L+n-2)$ is the separation constant and is also an eigenvalue of the operator $\Lambda^2$ ($\ref{21}$).

For convenience, we set a new variable $x = \omega r^2$ and function $R (x) = x^a e^{-x/2} g(x)$  and substitute them into the Eq. ($\ref{20}$). We obtain a confluent hypergeometric equation for $g(x)$, which will give a solution  of Eq. ($\ref{20}$) in terms of a special function
\begin{eqnarray}
R_{N L}(r) = C_{NL}  r^{L'} e^{-\frac{ \omega r^2}{2}} {}_1F_1 (-N; L'+\frac{n}{2}; \omega r^2) \label{22}
\end{eqnarray}
where
$L'(L'+n-2) =L(L+n-2) + 2 c_{a}.$

Hence we have the discrete energy $E_a$ of $H_a$ from (\ref{22}) as
\begin{eqnarray}
E_a=2\omega\left(N_a+\frac{L_a'}{2}+\frac {n}{4} \right),\label{23}
\end{eqnarray}
Thus, the energy spectrum of the our $N$-dimensional double singular oscillators
\begin{eqnarray}
\label{24}
4Z= E_1 + E_2 =2 \omega\left(N_1+N_2+\frac{L'_1+L'_2}{2}+\frac {n}{2}\right),
\end{eqnarray}

\subsection{Variables separation  in spherical coordinates}

The Schr\"{o}dinger equation for the generalized MICZ-Kepler system with non central terms  has the next form ($\ref{17}$):
  %================================================================
\begin{equation}
\eqalign{
%\begin{eqnarray}
%\hat
 {H'} \ \Psi({\bf r})
\ =
\left[
%\hat 
{ H'_{0}}+
\frac{\lambda_1}{r(r+x_{n+1})} +
\frac{\lambda_2}{r(r-x_{n+1})}
\right)\Psi  =  \nonumber\\
 \qquad \qquad \equiv
  \left(-\frac{1}{2}
\frac{{\partial}^2}{\partial x_{\lambda} \partial x_{\lambda}}
-\frac{Z}{r} +
\frac{\lambda_1}{r(r+x_{n+1})} +
\frac{\lambda_2}{r(r-x_{n+1})}
\right) \Psi =
\nonumber\\
 \qquad \qquad \equiv
  \left(-\frac{1}{2}
\Delta
-\frac{Z}{r} +
\frac{\lambda_1}{r(r+x_{n+1})} +
\frac{\lambda_2}{r(r-x_{n+1})}
\right) \Psi =
E \psi
,\label{25} 
%\end{eqnarray}
}
\end{equation}

In the $(n+1)$-dimensional spherical coordinates
%=================================
\begin{equation}
\eqalign{
%\begin{eqnarray}
x_{n+1}=r\cos(\theta), \nonumber\\ 
x_{n} \quad =r\sin(\theta)\cos(\phi_{n-2}), \nonumber\\
\qquad ... \nonumber\\
\qquad ...\nonumber\\ 
x_{2} \quad =r \sin(\theta)\sin(\phi_{n-2})\cdots \sin(\phi_{1}) \cos(\phi_{0}), \nonumber\\
x_{1} \quad =r \sin(\theta)\sin(\phi_{n-2})\cdots \sin(\phi_{1}) \sin(\phi_{0}),
\label{26}
%\end{eqnarray}
}
\end{equation}
%=====================================================================
using the ansatz  $\Psi({\bf u, v})=\Psi(r, \theta, \phi_{n-2},\cdots,\phi_{0}, \phi'_{n-2},\cdots,\phi'_{0})=R(r) \Theta(\theta) \Phi(\phi, \phi')$.
the Schr\"{o}dinger equation of our generalized MICZ-Kepler system with non central terms rewritten in terms of separated equations of variables $r, \theta$ and angular variables $\phi, \phi'$
%==================================================
\begin{eqnarray}
\left[
\frac{1}{r^{n}}\frac{\partial}{\partial r} ( r^{n} \frac{\partial}{ \partial r})+
2\left(\frac{Z}{r}+E\right)-
\frac{\Lambda}{r^2}
\right]R (r)=0,\label{27} \\
\left[
\frac{1}{\sin^{n-1}\theta}\frac{\partial}{ \partial \theta}( \sin^{n-1}\theta\frac{\partial}{ \partial \theta})  -
\frac{\hat{L}^2+ 4 \lambda_{2}}{4 \sin^{2}\frac{\theta}{2}}-
\frac{\hat{L}^2 + 4 \lambda_{1}}{4 \cos^{2}\frac{\theta}{2}}+ \Lambda
\right]
\Theta(\theta)=0,\label{28} \\
\left[
\hat{L}^2 -L(L+n-2)
\right]
\Phi(\phi, \phi')=0,
\label{29}
\end{eqnarray}
%=============================================
where $\Lambda=\lambda (\lambda+n-1)$ is the separation constant and is also an eigenvalue of the operator ($\ref{28}$).
 Solutions of ($\ref{27}$) - ($\ref{28}$) are as follows 
\begin{eqnarray*}
R(r)=C_{k \lambda} r^{\lambda} e^{-\sqrt{-2 E}r} {}_1 F_1(-k + \lambda - Q/2, n+2 \lambda, 2 \sqrt{-2 E}r),
%\nonumber
\\
\Theta(\theta)=C_{\lambda J L}(1+\cos\theta)^{J'/2}(1-\cos\theta)^{L'/2} 
P_{\lambda - (J' + L')/2}^{( J'+\frac{n-2}{2}, L'+\frac{n-2}{2})}(\cos\theta),
%\nonumber\\
\end{eqnarray*}
where
\begin{eqnarray*}
J'(J'+n-2) =L(L+n-2)+ 4 \lambda_{1},\quad \\
L'(L'+n-2) =L(L+n-2) + 4 \lambda_{2}.
%\nonumber\\
\end{eqnarray*}
Hence, the energy spectrum
\begin{eqnarray}
E=-\frac{Z^{2}}{2 ( k+\frac{n+Q}{2})^{2}}=-\frac{Z^{2}}{2 ( n_{r} + n_{\theta} +\frac{ n+J' + L'}{2})^{2}}.\label{30}.\label{30}
\end{eqnarray}
coincides with {\cite {van5}  in appearance only. For full compliance, it is necessary to take into account in the definition of the principal quantum number 
$k =n_{r} + n_{\theta} + (J' + L'-Q)/2$ that $J', L' \to L$ when $\lambda_{1}=\lambda_{2}= 0$.

\subsection{Variables separation  in parabolic coordinates}

The Schr\"{o}dinger equation $H'\Psi=E\Psi$  ($\ref{25}$) is same, but the Laplace-Beltrami operator $\Delta$ in the hyperparabolic coordinates has the next form:
\begin{eqnarray*}
&&\Delta=
\frac{4}{u+v}
\left\{
\frac{1}{u^{\frac{n-2}{2}}}\frac{\partial}{\partial u} ( u^{\frac{n}{2}} \frac{\partial}{ \partial u})+
\frac{1}{v^{\frac{n-2}{2}}}\frac{\partial}{\partial r} ( v^{\frac{n}{2}} \frac{\partial}{ \partial v})
\right\}-
\frac{L^2}{u v}
\end{eqnarray*}
The parabolic coordinates  on the $S_{n-1}$ sphere are defined according to works  \cite{van5} - \cite{van6} by
\begin{equation}
\eqalign{
%\begin{eqnarray}
x_{n+1}=\frac{u-v}{2},\nonumber\\ 
x_{n} \quad =\sqrt{u v} \cos(\phi_{n-2}),\nonumber\\
\qquad ...\nonumber\\
\qquad ...\nonumber\\ 
x_{2} \quad =\sqrt{u v} \sin(\phi_{n-2})\cdots \cos( \phi_{0}),\nonumber\\ 
x_{1} \quad =\sqrt{u v} \sin(\phi_{n-2})\cdots \sin( \phi_{0}),\nonumber\\
r \ \quad = \frac{u +v}{2},\label{31}
%\end{eqnarray}
}
\end{equation}
where the $(n+1)$ $x_{\lambda}$ are Cartesian coordinates in the hyperparabolic coordinates, $\{\phi_{0},\dots, \phi_{n-2}\}$ are the hyperparabolic angles and the parabolic coordinates $u$, $v$ range from $0$ to $\infty$.
In order to separate variables, the wave function is now choiced as follows
 $\Psi({\bf u, v})=\Psi(u, v, \phi_{n-2},\cdots,\phi_{0}, \phi'_{n-2},\cdots,\phi'_{0})=U(u) V(v) \Phi(\phi, \phi')$.

Thus, the Schr\"{o}dinger equation of our generalized MICZ-Kepler system with non central terms in the hyperparabolic coordinates can be rewritten in terms of separated equations of variables $u, v$ and angular variables $\phi, \phi'$
%==================================================
\begin{eqnarray}
\left[
\frac{1}{u^{\frac{n-2}{2}}}\frac{\partial}{\partial u} ( u^{\frac{n}{2}} \frac{\partial}{ \partial u}) -
\frac{L (L+n-2)+4 \lambda_{1}}{u}+
\frac{Z+E u}{2}
-P
\right]U (u)=0,\label{32} \\
\left[
\frac{1}{v^{\frac{n-2}{2}}}\frac{\partial}{\partial r} ( v^{\frac{n}{2}} \frac{\partial}{ \partial v}) -
\frac{L (L+n-2)+4 \lambda_{2}}{v}+
\frac{Z+E v}{2}
+P
\right]V (v)=0,\label{33} \\
\left[
\hat{L}^2 -L(L+n-2)
\right]
\Phi(\phi, \phi')=0,
\label{34}
\end{eqnarray}
%=============================================
where $P$ is the separation constant.

Solutions of (\ref{32}) and (\ref{33}) are given by the confluent hypergeometric polynomials,
\begin{equation}
\eqalign{
%\begin{eqnarray}
U (u)=C_{n_{1} J} u^{J'/2} e^{-\frac{\sqrt{-2 E}u}{2} } {}_1 F_1(-n_{1}, \frac{n}{2}+J', \sqrt{-2 E}u),\nonumber\\
V (v)=C_{n_{2} L} v^{L'/2} e^{-\frac{\sqrt{-2 E}v}{2}} {}_1 F_1(-n_{2}, \frac{n}{2}+L', \sqrt{-2 E}v),
\label{35}
%\end{eqnarray}
}
\end{equation}
where
\begin{eqnarray*}
J'(J'+n-2) = L(L+n-2) +4 \lambda_{1};\quad
-n_{1}=\frac{J'}{2}+\frac{n}{4} +\frac{P-Z/2}{\sqrt{-2 E}};
\nonumber\\
L'(L'+n-2) =L(L+n-2) + 4 \lambda_{2};\quad
-n_{2}=\frac{L'}{2}+\frac{n}{4} + \frac{-P-Z/2}{\sqrt{-2 E}}.
\end{eqnarray*}
Set
\begin{eqnarray*}
n_{1}+n_{2}=\frac{Z}{\sqrt{-2 E}}- \frac{J'+L'+n}{2}.
\end{eqnarray*}
The energy spectrum is as follows
\begin{eqnarray}
\label{36}
E=-\frac{Z^{2}}{2 \{n_1+n_2+\frac{n+J'+L'}{2}\}^{2}}.
\end{eqnarray}
Making the identification $k= n_1+n_2+\frac{J'+L'-Q}{2}$, the energy spectrum becomes (\ref{30}).

\section{ Hidden symmetry algebra and overlap coefficients}

\subsection{($N \equiv 2n$)-dimensional oscillator }

In according to the previous section 2, in order to have the quadratic Hahn algebra QH(3) as a hidden symmetry  algebra, two conditions must be satisfied. 
First of all, the resulting Hamiltonian must be the sum of the two original ones. Secondly, each of the starting Hamiltonians must have a symmetry $SU(1,1)$.

In subsection 3.1 we have shown next two facts as detailed as possible:
\begin{itemize}
  \item the ($N \equiv 2n$)-dimensional  model of harmonic isotropic oscillator may be considered as the sum of the two  independent ones of dimension $n$;
%=================================
\begin{equation}
\eqalign{
%\begin{eqnarray}
\hat { H_{0}}  \psi({\bf u, v}) = Z \psi \equiv (H_1+H_2)  \psi = (Z_1+Z_2)  \psi \nonumber\\
\qquad  \equiv  \left[-\frac{1}{2}(
\frac{{\partial}^2}{\partial u_{i} \partial u_{i}} +
\frac{{\partial}^2}{\partial v_{i} \partial v_{i}}) +
V_{sho}(u_{i} u_{i}) +V_{sho}( v_{i} v_{i}) 
\right] \psi  =  \nonumber\\
\qquad  \equiv
%(H_1+H_2)  \psi = (Z_1+Z_2)  \psi=
\sum_{1\le a\le 2}H_a  \psi
=
\sum_{1\le a\le 2}
\left[
-\frac{1}{2}
\frac{\partial^2}{\partial x_{i} \partial x_{i}} + 
V_{sho}( x_{i} x_{i}) 
\right]   \psi   
\label{37}
%\end{eqnarray}
}
\end{equation}
where $u_i, v_i\,\,(i=1,\ldots, n)$ are Cartesian coordinates of the space
${\rm I \!R}^{n}$;  $Z_a $ are the eigenvalues of $H_a$, but $Z_1+Z_2=Z$ and
%=================================================EEEEEEEEEEEEEEEEEEE
the potential of a singular oscillator $V_{sho}$ is
%===================================================================
\begin{eqnarray}
V_{sho}( x_{i} x_{i}) =
V_{ho}( x_{i} x_{i}) +
\frac{c_a}{x_{i} x_{i}} =
\frac{ \omega^2  x_{i} x_{i}}{2}
+
\frac{c_a}{x_{i} x_{i}},
\label{38}
\end{eqnarray}

  \item   the $nD$ model of  singular oscillator is an exactly solvable model in the hyperspherical coordinates ($\ref{19}$) with the following solution  of Eq. ($\ref{22}$).
\end{itemize}

Fixing quantum numbers (for the angular part in our case) allows us to reduce the problem to one-dimensional. This in turn means that the symmetry of the problem will now be determined by the first equation ($ \ref {20} $), which is the radial part of the $n$-dimensional non-relativistic Hamiltonian
harmonic oscillator $ H_ {ho} $.
Carrying out the replacement
$ R (r) = r ^{\frac {1-n}{2}} u (r) $
and with some reassignment of the constants we have a convenient form of this equation
to compare with the results of \cite {Zhe1}, namely:
\begin{eqnarray}
H_{ho}=
\left[
-\frac{1}{4 \omega}
\frac{\partial^2}{\partial r^2} +
\frac{ \omega}{4}
r^2
+\frac{g}{2 \omega r^2}
\right] 
\label{39}
\end{eqnarray}
where $g=\frac{4 L(L+n-2) + 8 c_{a} + (n-3)(n-1)}{8}$.

The following realization of the $SU(1,1)$ operators solves our problem:
\begin{equation}
\eqalign{
%\begin{eqnarray}
J_0 \equiv H_{ho}; \qquad  \qquad  
J_2=
\frac{i}{2}
\left[r
\frac{\partial}{\partial r} +
\frac{1}{2}
\right]; 
\nonumber\\
J_1=
\frac{\omega r^2}{2} -J_0; \qquad
J_{\pm}=J_1 \pm iJ_2;
\label{40}
%\end{eqnarray}
}
\end{equation}
Thus, the described realization allows to represent the original Hamiltonian as a sum of two items, each of them has the desired symmetry algebra $SU(1,1)$ in a radial variable. The Casimir operator is $Q=\frac{8g-3}{16}$, and the angular part is fixed by its well-known quantum numbers.

It should be noted that in this analyzed case, the division of the original  real space ${\rm I \!R}^{N}$  space into 2 equal $n$D parts, as well as the reduction to one-dimensional consideration is not significant. 
To show this fact, we will give one more realization of the $SU(1,1)$ operators for any value of the dimension of the space ${\rm I \!R}^{n'}$  (that is, the repeated index $m$ is summed up from $1$ to $n'$):
\begin{equation}
\eqalign{
%\begin{eqnarray}
J_0 =
\left[
-\frac{1}{4 \omega }
\frac{\partial^2}{\partial x_{m} \partial x_{m}} + 
%V( u_{s} u_{s})
\frac{ \omega}{4}
 x_{m}  x_{m}
+\frac{g}{2 \omega  x_{m}  x_{m}}
\right];  
\quad 
J_1=
\frac{\omega}{2}x_{m}  x_{m} -J_0; 
\nonumber\\
J_2=
\frac{i}{2}
\left[x_{m} 
\frac{\partial}{\partial x_{m}} +
\frac{n'}{2}
\right]; 
\quad 
J_{\pm}=J_1 \pm iJ_2;
\quad  
 Q=\frac{n(n-4)+8g}{16}-\hat{L}^2.
\label{41}
%\end{eqnarray}
}
\end{equation}
This result is interesting in several aspects. It turns out that the quadratic Hahn algebra QH(3) as a hidden symmetry remains unchanged with different decomposition of the original real space ${\rm I \!R}^{N}$ into components and also  when it's generalized to the model of a singular oscillator.

\subsection{ $(n+1)D$ related (MICZ-)Kepler-like systems}

 In according to  ($ \ref {16} $),  dual analog  of  so-called a double singular oscillator is  the generalized MICZ-Kepler system with non central terms. In the hyperparabolic coordinates ($ \ref {31} $) it can be rewritten in terms of separated equations of variables $u, v$ ($ \ref {32} $) - ($ \ref {33} $) and angular variables $\phi, \phi'$ ($ \ref {34} $). It's solutions for (\ref{32}) and (\ref{33}) are given by the confluent hypergeometric polynomials  ($ \ref {35} $) in subsection 3.3.

In other words, we have a similar situation as with a HO. The original Hamiltonian can be represent as a sum of two almost  identical  item in the parabolic coordinates $u$, $v$ respectively (Eqs. (\ref{32}) - (\ref{33})), i.e. if we fix the quantum numbers of the angular part we can reduce the problem to one-dimensional $u(v)$.  The latter
and will determine the symmetry of the problem now.
The following realization of the $SU(1,1)$ operators solves our problem in this case:
\begin{equation}
\eqalign{
%\begin{eqnarray}
J_0 =
\frac{1}{2 \gamma }
\left[
\frac{1}{x^{\frac{n}{2}-1}}\frac{\partial}{\partial x} ( x^{\frac{n}{2}} \frac{\partial}{ \partial x}) -
\frac{c_a}{x} 
% - \gamma^2 x
+ \frac{E}{2}x
\right]; 
\quad  
J_1=
\gamma x -J_0;
\nonumber\\
J_2=
i
\left[x
\frac{\partial}{\partial x} +
\frac{n}{4}
\right]; 
J_{\pm}=J_1 \pm iJ_2;
\quad  
 Q=\frac{n(n-4)}{16}+c_a.
\label{42}
%\end{eqnarray}
}
\end{equation}
where are $c_a=L (L+n-2)+4 \lambda_{a};  \ E=-2  {\gamma}^2$.

Thus, the described realization allows to represent the original Hamiltonian of $(n+1)D$ related MICZ-Kepler systems as a sum of two items, each of them  has the desired symmetry algebra $SU(1,1)$. The angular part is fixed by its well-known quantum numbers.

\subsection{The overlap coefficients}
Above we  have shown that the dual models to  the generalized version of the
KS-transformation models  ($N\equiv2n$)-dimensional singular oscillator and
the $(n+1)$-dimensional generalized MICZ-Kepler system)  have the exact analytical solutions in according to hidden simmetry as  the quadratic Hahn algebra QH(3).
The solvability of the Schr\"{o}dinger equation of the these models by the variables separation method in spherical and parabolic (cylindrical) coordinates connects with $SU(1,1) \oplus SU(1,1)$ dynamical symmetry. So, in according to the work \cite{Zhe1} the overlap coefficients between wavefunctions in these coordinates are Clebsch-Gordan coefficients (CGC) for $SU(1,1)$ algebra.
 Thus, on the one hand, the diagonalization of the operator $K_1$ corresponds  to choosing the unconnected basis 
$|n_1, \lambda_{1}> \otimes |n_{2}, \lambda_{2}>$ in the space of the direct sum $SU(1,1) \oplus SU(1,1)$ and, on the other hand, to separation of the variables in parabolic (cylindrical) coordinates. 
Similarly, on the one hand, the diagonalization of the operator $K_2$ corresponds to choosing the connected basis 
$|n_{12}, \lambda_{12}>$ in the space of the direct sum $SU(1,1) \oplus SU(1,1)$ and, on the other hand, to separation of the variables in spherical coordinates. 
The overlaps between eigenstates of the operations $K_1$ and $K_2$ (or the wavefunctions in these coordinate systems) can be written in terms of CGC for $SU(1,1)$.
An explicit it's expression for CGC in terms of Hahn polinomials can be found in the work \cite{Zhe1}:
\begin{eqnarray}
C_{n, \lambda_{1}; N - n, \lambda_{2}}^{j,\lambda_{12}}=
h_n w_{p} \ {}_{3}F_{2}
\left[
\begin{array}
  [c]{ll}%
-n, \ -p, \ 2\lambda_{1} + \lambda_{2} +p -1&
   \\
&; 1
  \\
\quad \quad \quad  -N, \quad \quad \quad 2\lambda_{1}&
  \end{array}
\right]
\label{43}
\end{eqnarray}

\section{The Higgs algebra as a commutant in $\mathcal{U}(U(2n))$}\label{sec_Higgs}

In this section we would like to explore hidden symmetry algebra as Higgs/Hahn algebras in the context of the new scientific direction, which has recently become particularly fashionable - to obtain a Howe duality setting for the interpretation of the Askey – Wilson (AW) and its degenerate algebras as commutants \cite{Zhe13}, \cite{Zhe25}-\cite{Rowe28}.
Therefore, the goal of this section is to discuss the Howe duality of Higgs-Hahn algebra for $(N \equiv 2n)D$ harmonic oscillator.

By definition, an $m$-dimensional orthogonal group $O(m)$ is a collection of all linear transformations in $m$-dimensional linear space, which leave invariant the sum of squares of the components of any vector $x=(x_{\alpha}), \ \alpha=1,\dots, m$ from this space $ x^2=x_{\alpha} x_{\alpha}=x_1^2+\dots+x_m^2$.

Choose a subalgebra $ O(n) \oplus O(n) $ in $ U(2n) $ generated by all rotations $L_a \equiv L_{ij}$, where $a \in \left( i, j=1, \dots, n \right)$ and $L_b \equiv L_{ij}$, where $b \in (i, j=n+1, \dots, 2n)$ that leave the norm $x_a^2=x_1^2+\dots+x_n^2$ and $x_b^2=x_{n+1}^2+\dots+x_{2n}^2$ constant correspondingly; obviously $ [L_a, L_b] = 0 $.
Define the next 3 operators
\begin{equation}
\eqalign{
%\begin{eqnarray}
\label{44}
 A_+=({a_1^{\dagger}}{}^{2}+ \dots +{a_n^{\dagger}}{}^{2})({a_{n+1}}^{2}+\dots +{a_{2n}}^{2}) \nonumber\\
 A_-=({a_1}^{2}+ \dots +{a_n}^{2})({a_{n+1}^{\dagger}}{}^{2}+ \dots +{a_{2n}^{\dagger}}{}^{2})   \nonumber\\
% D&=(N_1+N_2+N_3+N_4+N_5+N_6+N_7+N_8)-(N_{9}+N_{10}+N_{11}+N_{12}+N_{13}+N_{14}+N_{15}+N_{16}).
% D&=\sum_{1\le a\le 8} N_a - \sum_{9\le b\le 16} N_b.
D= \sum\limits_{a=1}^{a=n}  N_a  - \sum\limits_{b=n+1}^{b=2n} N_b.
% D&=\sum_{a} N_a - \sum_{b} N_b.
%\end{eqnarray}
}
\end{equation}
$A_\pm$ and $D$ are manifestly invariant under the rotations generated by $L_{a}$ and $L_{b}$ and they clearly commute with $H$ [thus belonging to $U(2n))$]. All other elements of the commutant are built from those.
It is easy to get the following formulas or commutators:
\begin{eqnarray*}
 [D,A_\pm]=\pm 4A_\pm,
\end{eqnarray*}
as well as the identities \cite{Zhe13}:
\begin{equation}
\eqalign{
%\begin{eqnarray}
\label{45}
{a_i^{\dagger}}{}^{2}{a_i}^{2}=({N_i}-1)N_i \nonumber\\
{a_i}^{2}{a_i^{\dagger}}{}^{2}=({N_i}+1)(N_i+2) \nonumber\\
{a_i}^{2}{a_j^{\dagger}}{}^{2}+{a_i^{\dagger}}{}^{2}{a_j}^{2}=2N_iN_j+N_i+N_j-4{L_{ij}}^{2},\qquad i,j=1,\dots,2n.
%\end{eqnarray}
}
\end{equation}
Find the remaining commutator $ [A_{+}, A_{-}]$:
\begin{eqnarray*}
[A_{+}, A_{-}]= 4 (\sum_{i=1}^{i=n} \sum_{j=1}^{j=n}{a_i^{\dagger}}{}^{2} {a_j}^{2})(  \sum_{b=n+1}^{b=2n}  N_b  + n/2) -4 ( \sum_{a=1}^{a=n}  N_a  + n/2) (\sum_{i=n+1}^{i=2n} \sum_{j=n+1}^{j=2n} {a_i^{\dagger}}{}^{2} {a_j}^{2})
\end{eqnarray*}
It with the help of  (\ref{45}) is readily converted to
\begin{eqnarray*}
[A_{+}, A_{-}]= 4 ((\sum_{i=1}^{i=n}  N_i)^2 +(n-2)\sum_{i=1}^{i=n}  N_i - 4 \sum_{(i<j)=1}^{(i<j)=n} L_{ij}^2)( \sum_{b=n+1}^{b=2n}  N_b  + n/2) \\
- 4 (\sum_{a=1}^{a=n}  N_a + n/2)((\sum_{i=n+1}^{i=2n}  N_i)^2 +(n-2) \sum_{i=n+1}^{i=2n}  N_i - 4 \sum_{(i<j)=n+1}^{(i<j)=2n} L_{ij}^2)
\end{eqnarray*}

Since
\begin{eqnarray*}
\sum_{a=1}^{a=n}  N_a=\frac{1}{2}(\frac{H}{\omega}+D-n), \qquad \sum_{b=n+1}^{b=2n}  N_b=\frac{1}{2}(\frac{H}{\omega}-D-n),
\end{eqnarray*}
upon substituting and after some  transformations, one obtains
\begin{equation}
\eqalign{
%\begin{eqnarray}
\label{46}
 [A_+,A_-]=\left[(\frac{H}{\omega})^{2} + 8\left(\sum\limits_{(i<j)=1}^{(i<j)=n} L_{ij}^2 + \sum\limits_{(i<j)=n+1}^{(i<j)=2n} L_{ij}^2\right)  + n(n-4)\right]D \nonumber\\
\qquad \qquad -8 \left( \sum\limits_{(i<j)=1}^{(i<j)=n} L_{ij}^2 -
\sum\limits_{(i<j)=n+1}^{(i<j)=2n} L_{ij}^2 \right) \frac{H}{\omega} -D^{3}
%\end{eqnarray}
}
\end{equation}
Thus, we have a Higgs algebra with structural constants of algebra, given as
\begin{equation}
\eqalign{
%\begin{eqnarray}
\label{47}
 \alpha_1=(\frac{H}{\omega})^{2}+8\left(\sum_{(i<j)=1}^{(i<j)=n} L_{ij}^2 +  \sum_{(i<j)=n+1}^{(i<j)=2n} L_{ij}^2 \right) + n(n-4) \nonumber\\
 \alpha_2= -8 \left( \sum_{(i<j)=1}^{(i<j)=n} L_{ij}^2 -
\sum_{(i<j)=n+1}^{(i<j)=2n} L_{ij}^2 \right) \frac{H}{\omega}
%\end{eqnarray}
}
\end{equation}
Substituting  (\ref{44}) into  (\ref{11}) using the identities  (\ref{45}) and the resulting expression for the structure constant in  (\ref{47}) gives the following expressions for the operators
\begin{equation}
\eqalign{
%\begin{eqnarray}
\label{48}
 K_1=\frac{1}{2}\left[ \sum_{a=1}^{a=n}  N_a  - \sum_{b=n+1}^{b=2n} N_b\right]  \nonumber\\
% K_2&={L_{12}}^{2}+{L_{13}}^{2}+{L_{14}}^{2}+{L_{23}}^{2}+{L_{24}}^{2}+{L_{34}}^{2}\\
 K_2=\sum_{(i<j)=1}^{(i<j)=2n} L_{ij}^2+\frac{n(n-2)}{4}   \nonumber\\
 K_3=[K_1,K_2],
%\end{eqnarray}
}
\end{equation}
These operators obey the commutation relations of the Hahn algebra
with structure constants given as
\begin{equation}
\eqalign{
%\begin{eqnarray}
\label{49}
 \delta_1=-\frac {\alpha_2}{4} = 2  \left( \sum_{(i<j)=1}^{(i<j)=n} L_{ij}^2 -
\sum_{(i<j)=n+1}^{(i<j)=2n} L_{ij}^2 \right) \frac{H}{\omega} \nonumber\\
 \delta_2=\phantom{-}\frac{\alpha_1}{2}=\frac{1}{2}(\frac{H}{\omega})^{2}+4\left(\sum_{(i<j)=1}^{(i<j)=n} L_{ij}^2 +  \sum_{(i<j)=n+1}^{(i<j)=2n} L_{ij}^2 + \frac{n(n-4)}{8}  \right)
%\end{eqnarray}
}
\end{equation}

\subsection{The Howe duality connection}

Now we can consider the duality of Howe. It is known \cite{Howe27} - \cite{Rowe28} that the state space of an $2n$-dimensional harmonic oscillator is realized both in the representation $SU(1,1)$ and $U(2n)$. Taking this fact into account, it can be shown that the embedding of a Hahn algebra into the double tensor product of one pair of $SU(1,1)$ is in duality with the commutator $ O(n) \oplus O(n) $ in the universal algebra of another algebra of the pair $U(2n)$.

For this purpose, we consider the addition of $2n$ metaplectic representations (\ref{6}), grouped into two pairs.
It means, that  if we take
\begin{eqnarray*}
J^{(1-2n)}=J^{(1-n)}+J^{((n+1)-(2n))}
\end{eqnarray*}
where
\begin{equation}
\eqalign{
%\begin{eqnarray}
\label{50}
J_0^{(1-2n)}&=J_0^{(1-n)}+J_0^{((n+1)-(2n))} \equiv
\frac{1}{2}
\left[
(\sum\limits_{a=1}^{a=n}  N_a  + n/2) +
(\sum\limits_{b=n+1}^{b=2n} N_b+n/2)\right]  \nonumber\\
J_+^{(1-2n)}&=J_+^{(1-n)}+J_+^{((n+1)-(2n))} \equiv
\frac{1}{2}\left[
\sum\limits_{a=1}^{a=n}  a_a^{\dagger}{}^{2}+
\sum\limits_{b=n+1}^{b=2n} a_b^{\dagger}{}^{2}
\right]  \nonumber\\
J_-^{(1-2n)}&=J_-^{(1-n)}+J_-^{((n+1)-(2n))} \equiv
\frac{1}{2}\left[
\sum\limits_{a=1}^{a=n}  a_a^{2}+\sum\limits_{b=n+1}^{b=2n} a_b^{2}\right]
%\end{eqnarray}
}
\end{equation}
then in accordance with the implementation of the operators of the Hahn algebra as in (\ref{8})  we get

\begin{eqnarray*}
K_1=J_{0}^{(1)}-J_{0}^{(2)} \equiv J_{0}^{(1-n)}-J_{0}^{((n+1)-(2n))}=\frac{1}{2}
\left[ \sum_{a=1}^{a=n}  N_a  - \sum_{b=n+1}^{b=2n} N_b\right]
\end{eqnarray*}
which coincides with the result in the the commutant approach given in (\ref{48}).

Similarly for $K_2$ we will have
\begin{eqnarray*}
K_2=C^{(1-2n)}=\left[J_0^{(1-n)}+J_0^{((n+1)-(2n))}\right]^{2}-\left(J_0^{(1-n)}+J_0^{((n+1)-(2n))}\right)  \nonumber\\
\qquad  \qquad  \qquad  -\left(J_+^{(1-n)}+J_+^{((n+1)-(2n))}\right)\left(J_-^{(1-n)}+J_-^{((n+1)-(2n))}\right)
\end{eqnarray*}

or taking into account (\ref{50}) 
\begin{eqnarray*}
K_2=\frac{1}{4}(\frac{H}{\omega})^{2}-\frac{1}{2}\frac{H}{\omega}-\frac{1}{4}
\left(
\sum\limits_{a=1}^{a=n}  a_a^{\dagger}{}^{2}+
\sum\limits_{b=n+1}^{b=2n} a_b^{\dagger}{}^{2}
\right)
\left(\sum\limits_{a=1}^{a=n}  a_a^{2}+
\sum\limits_{b=n+1}^{b=2n} a_b^{2}\right)
\end{eqnarray*}
Given the identities (\ref{45}) , we obtain
\begin{eqnarray*}
K_2=\frac{1}{4}(\frac{H}{\omega})^{2}-\frac{1}{2}\frac{H}{\omega}-\frac{1}{4}
\left[
A_++A_-
+
\left(
\sum_{a=1}^{a=n}
N_a
\right)^2
+
\left(
\sum_{b=n+1}^{b=2n}
N_b
\right)^2
\right]
-   \\
\qquad  \ - \frac{(n-2)}{4}
\left(
\sum_{a=1}^{a=n}
N_a
+
\sum_{b=n+1}^{b=2n}
N_b
\right)
+\sum_{(i<j)=1}^{(i<j)=n} L_{ij}^2 +
\sum_{(i<j)=n+1}^{(i<j)=2n} L_{ij}^2
\end{eqnarray*}
or
\begin{equation*}
\eqalign{
%\begin{eqnarray}
K_2=-\frac{1}{8}
\left[
2(A_++A_-)
+D^2 - (\frac{H}{\omega})^{2} - n(n-4)
\right]
+
\sum_{(i<j)=1}^{(i<j)=n} L_{ij}^2 +
\sum_{(i<j)=n+1}^{(i<j)=2n} L_{ij}^2
%\end{eqnarray}
}
\end{equation*}
This expression is identical $K_2=\sum\limits_{(i<j)=1}^{(i<j)=2n} L_{ij}^2+\frac{n(n-2)}{4} $ in formula  (\ref{48}), that was obtained earlier in the search for operators commuting during rotations $L_{1-n}$ and $L_{((n+1)-(2n))}$.
All of the above explicitly states that the operator $K_2$, being as Casimir operator $SU(1,1)$, belongs to the commutant of $L_{1-n}$ and $L_{((n+1)-(2n))}$ in $\mathcal{U}(U(2n))$.

A similar calculation shows that the $SU(1,1)$ Casimir operator for the representation  $J^{(ij)}$ is given by the square of the corresponding rotation generator in $U(2n)$, namely
\begin{eqnarray}\label{51}
 C^{(ij)}=L_{ij}^{2}+\frac{n(n-4)}{16} .
\end{eqnarray}
It follows that the structural constants of the Hahn algebra become on the basis of (\ref{12}):
\begin{equation}
\eqalign{
%\begin{eqnarray}
\label{52}
 \delta_1=4\left(J_0^{(1-n)}+J_0^{((n+1)-(2n))}\right)\left(C^{(1-n)}-C^{((n+1)-(2n))}\right) \nonumber\\
\quad   \equiv
 2\frac{H}{\omega}
 \left[
 \sum\limits_{(i<j)=1}^{(i<j)=n} L_{ij}^2 -
\sum\limits_{(i<j)=n+1}^{(i<j)=2n} L_{ij}^2
\right]
;
\nonumber\\
 \delta_2=2\left(J_0^{(1-n)}+J_0^{((n+1)-(2n))}\right)^{2}+4\left(C^{(1-8)}-C^{((n+1)-(2n))}\right)  \nonumber\\
\quad  =\frac{1}{2}(\frac{H}{\omega})^{2}+4\left(
 \sum\limits_{(i<j)=1}^{(i<j)=n} L_{ij}^2 +
\sum\limits_{(i<j)=n+1}^{(i<j)=2n} L_{ij}^2
 +\frac{n(n-4)}{8} \right)
%\end{eqnarray}
}
\end{equation}
in perfect correspondance with  (\ref{49}). 

Thus, it is established that the embedding of a Hahn algebra in $SU(1,1) \otimes  SU(1,1)$ leads to its description as a commutant in $\mathcal{U}(U(2n))$ in according to the pairing of representations $SU(1,1)$ and $U(2n)$ in the framework of Howe's duality.

\section{Dimensional reduction to the singular oscillator}
To carry the dimensional reduction of the $2n$-dimensional IHO in the approach with the 
embedding of the discrete version of the Hahn algebra in a double tensor product $SU(1,1) \otimes  SU(1,1)$ it is necessary to pay attention to the structural constants $\delta_1$ and $\delta_2$ in (\ref{12}).
In according to formula (\ref{51}), the Casimir operators are included in its definition  are grand angular momentum
operators $\hat L_i^{2},\ i=1,2$ in $n$-dimensional hyperspherical coordinates ($\phi_1, \dots, \phi_{n-1}$ are the hyperspherical angles and $r$ is the hyperradius). 
Therefore, eliminating the angular parts (hyperspherical angles)
by separating the variables and performing the 
 gauge transformation $\mathcal{O} \to\widetilde{\mathcal{O}}=r_i^{(n-1)/2}\mathcal{O} r_i^{-(n-1)/2}$ 
for the radial part, we actually obtain the Hamiltonian of the singular oscillator in two dimensions:
\begin{equation}
\label{53}
\eqalign{
%\begin{eqnarray}
 \widetilde{H}=2 \omega \left[\widetilde{J}_0^{(1-n)}+\widetilde{J}_0^{((n+1)-(2n))}\right] \nonumber\\
 \quad =\frac{1}{2}\left[
 -\left(\frac{\partial^{2}}{\partial r_1^{2}}+\frac{\partial^{2}}{\partial r_2^{2}}\right)+\omega^2 \left( r_1^{2}+r_2^{2}\right) -\frac{a_1}{r_1^{2}}-\frac{a_2}{r_2^{2}}
 \right],
%\end{eqnarray}
}
\end{equation}
where $a_i=\hat L_i^{2}+\frac{(n-3)(n-1)}{4},\ i=1,2.$
 
 However, do not forget that each dimension $i$ corresponds to it's $n$-dimensional hyper-radius $r_i$.
 The motion constants are obvious and equal to the operators of the Hahn algebra $ K_i $, which is here the algebra of hidden symmetry. They can be 
easyly finding in  according to the above material.

\section{Our dual models in  the class of quasi-exactly systems}

In the previous sections we are considered the dual connection with $N$-dimensional singular  oscillator and the general MICZ-Kepler system in  the class of exactly solvablle problems.
On the other hand, recall, that there are quasi-exactly problems which occupy an intermediate place between exactly solvable problems and non-solvable ones. 
It is also well known that the theory of quasi-exact systems (CES) gives
the following generalization or family of potentials, which includes HO, in the direction of degrees $ r $ less than 2, for a fixed $ N $, by
\begin{equation}
\label{54}
 V_{<2}(r)\ =V_{sho} + a r + \frac{b}{r}\  =
 \ \frac {\omega^2 r^2}{2}  + \frac{c}{r^2} + a r + \frac{b}{r}\
\end{equation}
with the eigenfunctions 
\begin{equation}
\label{55}
        R (r) \ = \  p_{N-1} (r) r^{l'-c'}  e^{-\frac{b'}{2}\, r^2  - a'\, r}\ ,
\end{equation}
where there are the following reassignment of constants from work \cite{Turbiner1} to our designation $\omega^2=2 b'^2; \quad a =2 a' b';  \quad b=-a' (D - 2c');  \quad c=c'(c'-D+1); \quad d=a'^2 - b'(2 N+D -1-2c'); \quad D=d'+2l'-1$; \quad $p_{N-1}(r)$ are polynomial of the ($N-1$)-th degree.

This QES potential appears in a number of applications to the systems with two electrons (\cite{Turbiner2} -  \cite{Turbiner3}).

Also at present we want to go to oppositive direction and to consider the potential which includes HO and degrees of the variable more than 2. The QES theory gives the other generalization in this direction or the family of potentials for a fixed $N$, by
see \cite{Turbiner1},
\begin{equation}
\label{56}
      V_{>2}(r) =V_{sho} + b r^4 + a r^6   \ =
        (\omega r)^2/2 + \frac{c}{r^{2}} + b r^4 + a r^6   \ ,
\end{equation}
with the eigenfunctions 
\begin{equation}
\label{57}
        R (r) \ = \  p_{N-1} (r^2) r^{l'-c'}  e ^ {-\frac{a'r^4}{4} - \frac{b'r^2}{2}},
\end{equation}
where the following reassignment of constants from work \cite{Turbiner1}  to our designation $\omega^2=2[b'^2 - (4 N+D - 2c'-1)a']; \quad a = a'^2;  \quad b=2 a' b';  \quad c= c'(c' - D+1)$.

The one-dimensional Hamiltonian of the nonrelativistic quantum systems with this anharmonic potential $(\ref{56})$ is well known as the crucial example that is stimulated the investigation of quasi-exactly solvable systems.

Thus, we  can offer four different models of the $(N \equiv 2n)D$ anisotropic and nonlinear anharmonic oscillator in QES class. Each model is represented by a sum of two independent (this provides an anisotropic effect - $\omega_1 \neq \omega_2$ ) oscillators of dimensions $ D = n $   with their various nonlinear anharmonic terms of the potential. 
In other words, we will further consider the following Hamiltonians of dimension $ D = n $:
\begin{equation}
\label{58}
\eqalign{
%\begin{eqnarray}
_{<2}H &= \left[-\frac{1}{2}
\frac{\partial^2}{\partial x_{s} \partial x_{s}} +  V_{<2}( x_{s} x_{s}) \right];\nonumber\\
V_{<2}(x_{s} x_{s}) &=  V_{sho}( x_{s} x_{s}) + \frac{b}{ \sqrt{ x_{s} x_{s} }}
 + a \sqrt { x_{s} x_{s} }  \nonumber\\
& =  \frac {\omega^2  x_{s} x_{s}}{2}  + \frac{c}{ x_{s} x_{s}}  + \frac{b}{ \sqrt{ x_{s} x_{s} }}
 + a \sqrt { x_{s} x_{s} } \  
% \end{eqnarray}
}
\end{equation}
and
\begin{equation}
\label{59}
\eqalign{
%\begin{eqnarray}
_{>2}H &=\left[-\frac{1}{2}
\frac{\partial^2}{\partial x_{s} \partial x_{s}} + 
 V_{>2}( x_{s} x_{s})
\right];\nonumber \\
V_{>2}(x_{s} x_{s})\ &= V_{sho}( x_{s} x_{s}) +
 b (x_{s} x_{s})^{2} + a (x_{s} x_{s})^{3} \nonumber\\
  &= \ \frac {\omega^2  x_{s} x_{s}}{2}  + \frac{c}{ x_{s} x_{s}} + b (x_{s} x_{s})^{2} + a (x_{s} x_{s})^{3}. 
%\end{eqnarray}
}
\end{equation}
%===================sequentially as follows:

In hyperspherical coordinates, the potentials of these Hamiltonians are successively reduced to the potential ($ \ref{54} $) and ($ \ref{56}$).
In any case, the final wave function $  \psi({\bf u, v})$  of Eq. ($\ref{37}$) will be represented by the product of the wave functions of each oscillator $\Psi({\bf r_a}) \equiv  R (r_a) \Omega(\phi_a)$ of Eq. ($\ref{58}$) -  ($\ref{59}$)
Recall that the hyperradius part  $R (r_a)$ of the eigenfunction $\Psi({\bf r_a})$ has the form  ($ \ref{55}$)  and ($ \ref{57} $), respectively, depending on the potential
($ \ref{54}$) and ($ \ref{56} $).

Thus, now we are ready to determine the potential of the dual analog for
our generalizations of the $(N \equiv 2n)D$ oscillator [ 4 models  from the combination ($ \ref{58} - \ref{59} $) ] as $n+1$  related MICZ-Kepler systems in different coordinates. 
However, the analysis performed in \cite{lan0} showed the absence of new solvable models for the dual partner or the general MICZ-Kepler system in this generalization for the spherical coordinates. Therefore, it is not considered further and the solvability of the Schr\"{o}dinger equation of the  MICZ-Kepler problems by the variables separation method will be discussed only in QES class for parabolic coordinates.
In according to  its definition ($ \ref{31}$) and  using the identities  (\ref{16}) it is not hard to receive:
\begin{eqnarray*}
 \qquad r= \sqrt{ x_{\lambda} x_{\lambda}} = u_{s}  u_{s} + v_{s}  v_{s}; \
 \qquad  x_{n+1} = u_{s}  u_{s} - v_{s}  v_{s}\\
     \qquad 2 u_{s}  u_{s} = r + x_{n+1} \equiv u ; \ 
   \qquad 2  v_{s}  v_{s} = r - x_{n+1} \equiv v 
\end{eqnarray*}

Therefore, we have additional additive terms, but each of them now is dependent on only one parabolic variable. 
Thus, the Schr\"odinger equation of our generalized MIC-Kepler system with new terms in the hyperparabolic coordinates can be rewritten in terms of separated equations of variables $u, v$ and angular variables $\phi, \phi'$
%==================================================
\begin{eqnarray}
\left[
\frac{1}{u^{\frac{n-2}{2}}}\frac{\partial}{\partial u} ( u^{\frac{n}{2}} \frac{\partial}{ \partial u}) -
\frac{L (L+n-2)+4 \lambda_{1}}{u} -
V_{u}
+Z_1 
%\frac{Z+E u}{2}
-P
\right]U (u)=0,\label{60} \\
\left[
\frac{1}{v^{\frac{n-2}{2}}}\frac{\partial}{\partial r} ( v^{\frac{n}{2}} \frac{\partial}{ \partial v}) -
\frac{L (L+n-2)+4 \lambda_{2}}{v} -
V_{v}
+Z_2
%\frac{Z+E v}{2}
+P
\right]V (v)=0,\label{61} \\
\left[
\hat{L}^2 -L(L+n-2)
\right]
\Phi(\phi, \phi')=0,
\label{62}
\end{eqnarray}
%=============================================
where $P$ is the separation constant.

Let us specify the value of  $V_{u} \equiv  V(u_{s} u_{s}) $  and 
 $V_{v} \equiv  V(v_{s} v_{s})  $ 
for each model
\begin{itemize}
 \item  { 
Model 1
 \begin{equation}
\label{63}
\eqalign{
%\begin{eqnarray}
%\label{37}
H \psi_1({\bf u, v}) &=  \left[{}_{<2}H_1 + {}_{<2}H_2\right]  \psi_{<2}({\bf u}) \psi_{<2}({\bf v})         \equiv    Z \psi_1({\bf u, v})  
\\
 \qquad V_{u} & = 
  \frac {- u E_1}{2}    + b_1 \sqrt{ \frac {2}{u} }
 + a_1 \sqrt { \frac {u}{2}  }
\nonumber\\
 \qquad V_{v} &= 
\frac {- v E_2}{2}    + b_2 \sqrt{ \frac {2}{v} }
 + a_2 \sqrt { \frac {v}{2}  } 
    \nonumber
%    \end{eqnarray}
}
\end{equation}
}
 \item 
 {
 Model 2
  \begin{equation}
\label{64}
\eqalign{
%\begin{eqnarray}
%\label{38}
H \psi_2({\bf u, v}) &= \left[{}_{<2}H_1 + {}_{>2}H_2\right]  \psi_{<2}({\bf u}) \psi_{>2}({\bf v})   
\equiv  Z \Psi_{<2}({\bf r_1}) \Psi_{>2}({\bf r_2})
\\
 \qquad V_{u} & = 
   \frac {- u E_1}{2}    + b_1 \sqrt{ \frac {2}{u} }
 + a_1 \sqrt { \frac {u}{2}  }
 \nonumber\\
 \qquad V_{v} &= 
\frac {- u E_2}{2}  + 
  \frac {b_2 u^2}{4} + \frac {a_2 u^3}{8}
         \nonumber
%  \end{eqnarray}
}
\end{equation}
}
\item 
 {
 Model 3
  \begin{equation}
\label{65}
\eqalign{
%\begin{eqnarray}
%\label{39}
H \psi_3({\bf u, v})& = \left[{}_{>2}H_1 + {}_{<2}H_2\right]  \psi_{>2}({\bf u}) \psi_{<2}({\bf v})   
\equiv  Z \Psi_{>2}({\bf r_1}) \Psi_{<2}({\bf r_2})
\\
 \qquad V_{u} &= 
 \frac {- u E_1}{2}  + 
  \frac {b_1 u^2}{4} + \frac {a_1 u^3}{8}
\nonumber\\
 \qquad V_{v} &=  
\frac {- v E_2}{2}    + b_2 \sqrt{ \frac {2}{v} }
 + a_2 \sqrt { \frac {v}{2}  } 
   \nonumber
%   \end{eqnarray}
}
\end{equation}
}
 \item 
 {
 Model 4
  \begin{equation}
\label{66}
\eqalign{
%\begin{eqnarray}
%\label{40}
H \psi_4({\bf u, v})& = \left[{}_{>2}H_1 + {}_{>2}H_2\right]  \psi_{>2}({\bf u}) \psi_{>2}({\bf v})  
\equiv  Z  \Psi_{>2}({\bf r_1}) \Psi_{>2}({\bf r_2})
\\
 \qquad V_{u}  &= 
\frac {- u E_1}{2}  + 
  \frac {b_1 u^2}{4} + \frac {a_1 u^3}{8} 
 \nonumber\\
 \qquad V_{v} &= 
\frac {- u E_2}{2}  + 
  \frac {b_2 u^2}{4} + \frac {a_2 u^3}{8}
 \nonumber
%       \end{eqnarray}
}
\end{equation}
}
\end{itemize}

At first glance, it seems that 4 models require solving 2 qualitatively different types of problems. 
In particular, for model 4 we will have the following equations after the 
 gauge transformation $\mathcal{O} \to\widetilde{\mathcal{O}}=x^{n/4}\mathcal{O} x^{-n/4}$ 
for $(\ref{60}) - (\ref{61})$:
\begin{equation}
\label{67}
\eqalign{
%\begin{eqnarray}
\left[
\frac{\partial^{2}}{\partial x^{2} } -
\frac{d_i }{x^2} -
\frac{a_i}{8} x^{2}+
\frac{E_i}{2} +
\frac{Z_i \pm P}{x}  -
\frac{b_i}{4} x 
\right]X (x)=0,
%\end{eqnarray}
}
\end{equation}
where $d_i=L (L+n-2)+4 \lambda_{1}-\frac{n(n-4)}{16},\ i=1,2.$

However, getting rid of irrationality by introducing a new variable $x \to z^2$ in one type of problem leads to the solution of a second type problem, which in turn coincides with the problem we considered earlier ($\ref{54}$)  with solution ($\ref{55}$).
In particular, for model 1 we will have the following equations after the 
 gauge transformation $\mathcal{O} \to\widetilde{\mathcal{O}}=z^{\frac{(n-1)}{2}}\mathcal{O} z^{-\frac{(n-1)}{2}}$ 
for $(\ref{60}) - (\ref{61})$:
\begin{equation}
\label{68}
\eqalign{
%\begin{eqnarray}
\left[
\frac{\partial^{2}}{\partial z^{2} } -
\frac{d_i }{z^2} +
2 E_i z^{2}+
4(Z_i \pm P) -
4 \frac{b_i}{z} -
4 a_i z \right]Z (z)=0,
%\end{eqnarray}
}
\end{equation}
where $d_i=4L (L+n-2)+16\lambda_{1}-\frac{(n-1)(n-3)}{4},\ i=1,2.$

Thus, we have obtained that in parabolic coordinates all proposed 4 models  
$(\ref{63}$) - ($\ref{66}$)   is the (quasi) exactly solvable models.

Now let us compare the results obtained with those available in the literature \cite{yeg2}. If we consider just a problem with potential $V_{ho}( x_{i} x_{i})  \equiv
\frac{ \omega^2  x_{i} x_{i}}{2}$ ($\ref{38}$), but without a condition  $\omega_1 = \omega_2$ , then we obtain in the $(N \equiv 2n)-D$  the sum of two independent harmonic oscillators that will be dual to the $n+1$ MICZ-problem with the potential $\cos \theta$:
\begin{eqnarray*}
%\frac {1} {r}
% \left[ 
\frac{ \omega_1^2 u_{s} u_{s} }{8}  + 
\frac{ \omega_2^2 v_{s} v_{s} }{8}  
% \right] 
& \equiv 
%\frac {1} {r}
% \left[ 
- E_1  u_{s} u_{s}  - E_2  v_{s} v_{s} 
% \right]  
\\
% \equiv 
\to 
\frac {1} {r}
\left[ 
- \frac {E_1 (r +x_{n+1})} {2} - \frac {E_2  (r - x_{n+1})} {2}   
\right] 
& \equiv 
% - E_1 (1 +\cos \theta) -  E_2 (1 - \cos \theta)  
- \frac {E_1} {2} (1 +\frac {x_{n+1}} {r}) - \frac {E_2} {2} (1 - \frac {x_{n+1}} {r})  \\
  = -
\frac {E_1+E_2} {2}
%(E_1+ E_2)  
-
%(E_1- E_2)   \cos \theta  \\
\frac {E_1-E_2} {2}  \cos \theta  
& \equiv -E_{MICZ} + \frac {\Delta w^2}{4} \cos \theta.   
 \end{eqnarray*}
 
The analog of the 4th order anisotropic potential term for oscillator system \cite{yeg2}  is in our  $(N \equiv 2n)-D$ case the sum of 2 harmonic oscillators with identical in value but different in sign potential coefficients for nonlinearity of the 4th order  ($b_1= - b_2$). In the $n+1$-dimensional space, this leads to the desired linear term:

\begin{eqnarray*}
%\frac {1} {r}
% \left[ 
b_1 ( u_{s} u_{s} )^2  + 
b_2 ( v_{s} v_{s} )^2  
% \right] 
& =
b
\left[ 
( u_{s} u_{s} )^2  -
 ( v_{s} v_{s} )^2 
 \right] 
  \\
\to 
\frac {b} {r}
\left[ 
\frac {(r +x_{n+1})^2} {4} - \frac {(r - x_{n+1})^2} {4}   
\right] 
& =
\frac {b} {4r}
\left[ 
4  \ r  \ x_{n+1}
\right] 
\\
=
b \ x_{n+1}
& \equiv b r \cos \theta.   
 \end{eqnarray*}

We note in particular that in our case there is still a new term with a higher degree of nonlinearity than the one considered above.

%\newpage

\section{Conclusion}
We have shown that $N$-dimensionall singular oscillator and  $(n+1)$-dimensional generalized MIC-Kepler system are dual to each other and the duality transformation is the generalized version of the Kustaanheimo-Stiefel transformation. 
The solvability of the Schr\"{o}dinger equation of the these problems by the variables separation method were given in different coordinates. We have successfully built the exact analytical solutions of the Schr\"{o}dinger equation for $N \equiv 2n$-dimensionall singular oscillator in double coordinates and  for the $(n+1)$-dimensional generalized MIC-Kepler system in spherical and parabolic coordinates.
The overlap coefficients between wavefunctions in these coordinates were shown to coincide with Clebsch-Gordan coefficients for $SU(1,1)$ algebra. 
It turns out that the quadratic Hahn algebra QH(3) as a hidden symmetry remains unchanged with different decomposition of the original space real space ${\rm I \!R}^{N}$ into components  in the framework of addition rule for $SU(1,1)$ algebra.
Also the  hidden symmetry algebra as Higgs/Hahn algebras was clearly shown by the commutant approach in the sense of Howe duality. 
A dimensional reduction was carried out to a singular oscillator of two dimensions, each variable  is the $n$-dimensional hyper-radius $r$.
The dual connection with $N \equiv 2n$-dimensional singular  oscillator and the $(n+1)$  general MICZ-Kepler system in  the class of  quasi-exact problems  was considered.
In the above framework of generalized KS transformations, 
some generalization of HO by  anisotropic  and nonlinear inharmonic terms and 
and its dual analog was shown and analyzed.
The exact analytical solutions of the Schrödinger equation for abovementioned  problems for QES class were discused and given for four series of dual quasi-exact solvable models.  
In particular, a comparison  with similar results in lower dimensions and its generalization were given.

\end{document}